\documentstyle[prd,aps,psfig]{revtex}
\def\be{\begin{equation}}
\def\ee{\end{equation}}
\def\ba{\begin{eqnarray}}
\def\ea{\end{eqnarray}}
\begin{document}
%
%\twocolumn
%
\draft
\title{The use of early data on $B \rightarrow \rho \pi$ decays}
\author{Helen R.\ Quinn and Jo\~ao P.\ Silva\footnote{Permanent address: 
	Instituto Superior de Engenharia de Lisboa,
	Rua Conselheiro Em\'{\i}dio Navarro,
	1900 Lisboa, Portugal.}}
\address{Stanford Linear Accelerator Center, Stanford University, Stanford,
	 CA 94309, USA}
\date{\today}
\maketitle
\begin{abstract}
This paper reviews the Dalitz plot analysis for the decays
$B^0(\bar B^0) \rightarrow \rho \pi \rightarrow \pi^+ \pi^- \pi^0$.
We discuss what can be
learned about the ten parameters in this analysis from untagged and
from tagged time-integrated data.
We find that,
with the important exception of the interesting $CP$ violating
quantity $\alpha$,
the parameters can be determined from this data sample -- and,
hence,
they can be measured at CLEO as well as at the asymmetric $B$ factories.
This suggests that the extraction of $\alpha$ from the time-dependent
data sample can be accomplished with a smaller data sample
(and, therefore, sooner) than would be required if all ten parameters
were to be obtained from that time-dependent data sample alone.
We also explore bounds on the shift of the true angle $\alpha$
from the angle measured from charged $\rho$ final states alone.
These may be obtained prior to measurements of the parameters
describing the neutral $\rho$ channel,
which are expected to be small.

\end{abstract}

%\narrowtext

\section{Introduction}
\label{sec:intro}

The $B$ factories at SLAC and KEK are now collecting data and
expect to produce  measurements of the $CP$ asymmetry in the mode
$B\rightarrow J/\psi K_S$ within one year,
measuring $\sin 2 \beta$. 
Preliminary results on this mode from CDF \cite{CDF}
already indicate that it is unlikely that a discrepancy with
the Standard Model will be found from this result alone.
This means that tests of the Standard Model mechanism for $CP$ violation will
rely upon our ability to measure further $CP$-violating parameters,
such as the angle $\alpha =\pi - \beta- \gamma$ of the Unitarity
triangle,
with sufficient accuracy to be sensitive to Standard 
Model relationships.
This will require that we master the removal of theoretical uncertainties
due to penguin diagrams in at least one of the available channels.
So far there are two sets of candidate decay modes,
$\pi \pi$ \cite{gronaulondon} and $\rho\pi$ \cite{SQ},
for which analyses to extract $\alpha$ using isospin relationships have
been suggested.
The first suffers from relatively small branching ratios\cite{cleopipi} 
and from the experimental difficulty of measuring the $\pi^0\pi^0$ 
branching ratio.
Some of the modes for the second case have recently been observed at
CLEO \cite{cleorhopi}.
These results are encouraging. 
They are close to model-dependent predictions and,
hence,
appear to reinforce estimates that the analysis suggested by
Snyder and Quinn will require of order 180 fb$^{-1}$,
or six years of running at BABAR design luminosity,
to complete \cite{BaBarbook}.

This paper revisits that analysis and reviews what intermediate steps can be 
made with earlier data samples.
In particular we stress that many of the parameters relevant to the
extraction of $\alpha$ can be predetermined from untagged,
and from tagged but time-integrated data samples.
The benefit of this is that,
once this is done,
only the $CP$-violating parameter $\alpha$ remains to be fit to the
full tagged and time-dependent Dalitz plot. 
Presumably it will then be possible to perform the $\rho \pi$ analysis
with smaller data samples than those needed if the full set of ten
parameters is fit to the tagged, time-dependent sample alone
(as has been done to date in Monte Carlo studies of this mode).
We also discuss bounds on the shift in $\alpha$ from penguin contributions.
These can be obtained,
by methods similar to those previously suggested for the $\pi\pi$ 
modes \cite{grossmanquinn,charles,pirjol},
even if the $\rho^0\pi^0$ amplitude is too small to 
be measured directly.
We will also discuss a bound which applies only in the $\rho \pi$
case,
having no parallel in the $\pi \pi$ case.

This paper presents a purely theoretical discussion,
with no simulations and no attempt to address issues of backgrounds
or of other modes that may contribute to the $\pi^+\pi^-\pi^0$
Dalitz plot \cite{sophieetc}.
Certainly these issues will be important.
The larger untagged data sample will also be the first place to 
explore the issues of further contributions to the Dalitz plot.
The approach of fixing as many parameters as possible from the untagged
and from the tagged but time-integrated data samples,
combined with improvements in machine luminosity,
such as those already under study for PEPII,
may make the extraction of a reliable value for $\alpha$ a reality
on a somewhat faster time scale than suggested by the estimates 
based on fits to tagged data only.

\section{Notation}

\subsection{Isospin decomposition}

We are interested in the decays from $B^+$ and $B^0$ into
$\rho \pi$ final states.
The decay amplitudes can be classified according to isospin:
$\{B^+, B^0 \}$ form an isospin doublet;
the final state $\rho \pi$ can have isospin $I_f=0$,
$I_f=1$,
and $I_f=2$ components.
In general,
the final state with $I_f=0$ can only be reached
with operators having $\Delta I = 1/2$;
the final state with $I_f=1$ can be reached
with operators having $\Delta I = 1/2$ or  $\Delta I = 3/2$;
and the final state with $I_f=2$ can be reached
with operators having $\Delta I = 3/2$ or  $\Delta I = 5/2$.
We denote the isospin amplitudes by $A_{\Delta I, {I_f}}$.
Thus,
\ba
a_{+0} & = &
a \left( B^+ \rightarrow \rho^+ \pi^0 \right)
=
\frac{1}{2} \sqrt{\frac{3}{2}} A_{3/2, 2}
- \frac{1}{2} \frac{1}{\sqrt{2}} A_{3/2, 1}
+ \frac{1}{\sqrt{2}} A_{1/2, 1}
\left[
- \frac{1}{\sqrt{6}}  A_{5/2, 2}
\right],
\nonumber\\
a_{0+} & = &
a \left( B^+ \rightarrow \rho^0 \pi^+ \right)
=
\frac{1}{2} \sqrt{\frac{3}{2}} A_{3/2, 2}
+ \frac{1}{2} \frac{1}{\sqrt{2}} A_{3/2, 1}
- \frac{1}{\sqrt{2}} A_{1/2, 1}
\left[
- \frac{1}{\sqrt{6}}  A_{5/2, 2}
\right],
\nonumber\\
a_{+-} & = &
a \left( B^0 \rightarrow \rho^+ \pi^- \right)
=
\frac{1}{2 \sqrt{3}} A_{3/2, 2}
+ \frac{1}{2} A_{3/2, 1}
+ \frac{1}{2} A_{1/2, 1}
+ \frac{1}{\sqrt{6}}  A_{1/2, 0}
\left[
+ \frac{1}{2 \sqrt{3}}  A_{5/2, 2}
\right],
\nonumber\\
a_{-+} & = &
a \left( B^0 \rightarrow \rho^- \pi^+ \right)
=
\frac{1}{2 \sqrt{3}} A_{3/2, 2}
- \frac{1}{2} A_{3/2, 1}
- \frac{1}{2} A_{1/2, 1}
+ \frac{1}{\sqrt{6}}  A_{1/2, 0}
\left[
+ \frac{1}{2 \sqrt{3}}  A_{5/2, 2}
\right],
\nonumber\\
a_{00} & = &
a \left( B^0 \rightarrow \rho^0 \pi^0 \right)
=
\frac{1}{\sqrt{3}} A_{3/2, 2}
- \frac{1}{\sqrt{6}}  A_{1/2, 0}
\left[
+ \frac{1}{\sqrt{3}}  A_{5/2, 2}
\right].
\label{eq:clebsch}
\ea
Similar relations hold for the $\bar B^0$ decay amplitudes,
which we denote by $\bar a_{-0}$,
$\bar a_{0-}$,
$\bar a_{-+}$,
$\bar a_{+-}$,
and $\bar a_{00}$,
respectively.
Note that our notation here is that $\bar a_{ij}$ is the amplitude for the
$\bar B^0$ to decay to a $\rho$ of charge $i$ and a $\pi$ of charge $j$.
The $CP$ relationships are thus
$CP (a_{ij}) = \bar a_{-i,-j}$. 
The $\bar B^0$ isospin components are
$\bar A_{\Delta I, I_f}$.

In the Standard Model,
there are tree-level amplitudes,
gluonic penguin amplitudes,
electroweak penguin amplitudes,
and final state rescattering effects.
The tree level $b \rightarrow u \bar u d$ decays have
both $\Delta I = 1/2$ and $\Delta I = 3/2$ components.
In contrast, the gluonic $b \rightarrow d$ penguins are pure
$\Delta I = 1/2$,
because the gluon is pure $I=0$.
Therefore,
the isospin amplitudes $A_{3/2,1}$ and $A_{3/2,2}$
only receive contributions (and, thus, weak phases) from
the tree-level diagrams. There are no diagrammatic $\Delta I=5/2$ contributions
at this order;
such effects arise only from electromagnetic corrections to the 
weak-decay diagrams.

A priori,
there is no hierarchy among the $\Delta I=1/2$ and $\Delta I =3/2$ 
isospin amplitudes. However,
the combination of isospin amplitudes involved in the decay
$B^0 \rightarrow \rho^0 \pi^0$ is generally argued to be  suppressed
because the tree-level and gluonic penguin diagrams can
only contribute to $B^0 \rightarrow \rho^0 \pi^0$ through a
color-suppressed recombination of the quarks in the final state.
This argument is not theoretically rigorous because
$B^0 \rightarrow \rho^0 \pi^0$ may be fed from other topologies
through strong final state rescattering.
Eventually experiment will tell us whether these effects are important.
 
In what follows we will neglect two contributions
which are suppressed in the SM.
The first contribution arises from the electroweak penguin diagrams,
which have the same weak phase structure as the QCD penguin diagrams,
but contribute to both $\Delta I = 1/2$ and $\Delta I = 3/2$
amplitudes.
As a result,
the electroweak penguin contributions are not removed by the
isospin-based analyses. 
However,
they are expected to be very small
in these channels \cite{BaBarbook,electroweak}.
The second contribution is due to a possible $\Delta I = 5/2$
isospin component,
included within square brackets in Eqs.~(\ref{eq:clebsch}).
In the SM,
this component comes from electromagnetic rescattering effects and,
thus,
it is suppressed by $\alpha \sim 1/127$.\footnote{Notice, however,
that this effect is important in $K \rightarrow \pi \pi$ decays
because there one has a strong hierarchy among the decay amplitudes,
$\mbox{Re} A_{2}/ \mbox{Re} A_{0} \sim 1/22$,
encoded into the $\Delta I = 1/2$ rule
\cite{DGH,BLS,electromagnetic}.}
Both effects could become relevant for $B^0 \rightarrow \rho^0 \pi^0$,
should this amplitude turn out to be very small. 

Henceforth,
we will only consider the tree-level and gluonic penguin diagrams.
These give contributions with weak phases
\ba
\frac{V_{ub}^\ast V_{ud}}{|V_{ub}^\ast V_{ud}|}
&=& e^{i \gamma} = - e^{-i(\beta + \alpha)},
\nonumber\\
\frac{V_{tb}^\ast V_{td}}{|V_{tb}^\ast V_{td}|}
&=& e^{- i \beta}.
\label{eq:ckm}
\ea
The first of these expressions is the weak phase of the
tree-level contributions.
Here we follow \cite{pdgbreview} and use the unitarity relationship 
to rewrite the penguin contribution as a dominant term proportional 
to the second of the CKM factors in Eq.~(\ref{eq:ckm})
plus a sub-dominant term proportional to the first CKM factor
in Eq.~(\ref{eq:ckm}).\footnote{The
term is sub-dominant in that it is a difference of up and charm
quark contributions and,
hence,
it vanishes in the limit that these two quark masses are taken to be equal.}
In what follows we always subsume this second term within the amplitudes
we refer to as $\Delta I=1/2$ tree amplitudes.
Since we do not calculate these quantities,
but rather discuss extracting them from fits to experiment,
this makes no difference to our analysis.
Note, however, that if the amplitudes extracted in this way are to be
compared to those calculated in any given model,
then the penguin contributions to our so-called ``tree''
amplitudes must be taken into account.
 
The interference between the two amplitude contributions 
in Eq.~(\ref{eq:ckm}) depends on the weak phase $\alpha$.
However,
we will show that,
as is usual with direct $CP$-violation effects,
any sensitivity to $\alpha$ is masked by an unknown
coefficient with large theoretical uncertainty.
Another weak phase
arises from the interference between the $B^0 - \overline{B^0}$ mixing,
$q/p = \exp[- 2 i (\beta - \theta_d)]$, and the tree-level diagrams.
For example,
\be
\frac{q}{p} \frac{\bar A_{3/2,2}}{A_{3/2,2}}
= e^{-2 i (\beta - \theta_d)} e^{- 2 i \gamma}
= e^{2 i (\alpha + \theta_d)},
\label{phase-measured}
\ee
where $\theta_d$ parametrizes a possible new physics contribution to
the phase in $B^0 - \overline{B^0}$ mixing.
In the SM,
$\theta_d=0$ and the two phases coincide;
other models have $\theta_d \neq 0$ and a difference arises.
The phase probed by the Snyder--Quinn method is $\alpha + \theta_d$
\cite{BLS}.
%%%%% I suspect there is an earlier reference where this is stated
%%%%% explicitly

\subsection{$B^0 \rightarrow \pi^+ \pi^- \pi^0$ decay amplitudes}

The $\rho$-mediated $B^0 \rightarrow \pi^+ \pi^- \pi^0$ decay amplitudes
may be written as
\ba
A = a \left( B^0 \rightarrow  \pi^+ \pi^- \pi^0 \right)
& = &
f_+ a_{+-} + f_- a_{-+} + f_0 a_{00},
\nonumber\\
\bar A = a \left( \overline{B^0} \rightarrow  \pi^+ \pi^- \pi^0 \right)
& = &
f_+ \bar a_{+-} + f_- \bar a_{-+} + f_0 \bar a_{00},
\label{decayamp:+-0}
\ea
where $f_\pm$ and $f_0$ are Breit-Wigner functions representing
$\rho^\pm$ and $\rho^0$, respectively,
and also include the $\cos{\theta}$ angular dependences
of the helicity zero $\rho$ decays.
The crucial observation made by Snyder and Quinn is that
the Breit-Wigner functions contain $CP$-even phases,
and that the interference between the different Breit-Wigner shapes
across the Dalitz plot provides experimental sensitivity to
the weak and strong phases contained in Eqs.~(\ref{decayamp:+-0}).
There is some systematic uncertainty associated with the 
specific choice made for the shape of the functions $f_\pm$ and $f_0$.
This uncertainty affects
primarily the corners of the Dalitz plot where the tails of two
such functions interfere;
these are the regions of the Dalitz plot from which one
extracts the interference between two distinct decay amplitudes.

It will prove convenient to rewrite  Eqs.~(\ref{decayamp:+-0}) as
\ba
A 
& = &
f_c a_{c} + f_d a_{d} + f_n a_n,
\nonumber\\
\bar A
& = &
f_c \bar a_{c} + f_d \bar a_{d} + f_n \bar a_n,
\label{decayamp:cdn}
\ea
where
\be
f_c = \frac{f_+ + f_-}{2},
\hspace{5ex}
f_d = \frac{f_+ - f_-}{2},
\hspace{5ex}
f_n = \frac{f_0}{2},
\ee
and
\ba
a_{c} = a_{+-} + a_{-+},
& \hspace{5ex} &
a_{d} = a_{+-} - a_{-+},
\hspace{5ex}
a_n = 2 a_{00},
\nonumber\\
\bar a_{c} = \bar a_{+-} + \bar a_{-+},
& \hspace{5ex} &
\bar a_{d} = \bar a_{+-} - \bar a_{-+},
\hspace{5ex}
\bar a_n = 2 \bar a_{00}.
\label{define:acd}
\ea
Notice that the $CP$-conjugate of $a_{d}$ is not
$\bar a_d$, but rather $- \bar a_d$.

As discussed in the appendix,
we parametrize the amplitudes $a_c$, $a_d$, and $a_n$ as
\ba
a_c
& = &
T e^{- i \alpha}
\left( 1 - z - r_0 \right),
\nonumber\\
a_d
& = &
T e^{- i \alpha}
\left( z_1 + r_1 \right),
\nonumber\\
a_n
& = &
T e^{- i \alpha}
\left( z + r_0 \right),
\label{master}
\ea
where $z$ and $z_1$ are $CP$-even, while $r_0$ and $r_1$
are $CP$-odd.
The quantity $T z_1$ contains the $CP$-even contributions to
the final state with $I_f = 1$,
summing the tree amplitude and the penguin amplitude
multiplied by cos$\alpha$.
$T r_1$ ($T r_0$) contains the $CP$-odd part,
given by the penguin contributions to
the final state with $I_f = 1$ ($I_f = 0$),
multiplied by $i \sin{\alpha}$.
Similarly,
the amplitudes contained in $q \bar A_f / p$ can be written as
\ba
\frac{q}{p} \bar a_c
& = &
e^{2 i \theta_d} T e^{i \alpha}
\left( 1 - z + r_0 \right),
\nonumber\\*[1mm]
\frac{q}{p} \bar a_d
& = &
e^{2 i \theta_d} T e^{i \alpha}
\left( - z_1 + r_1 \right),
\nonumber\\*[1mm]
\frac{q}{p} \bar a_n
& = &
e^{2 i \theta_d} T e^{i \alpha}
\left( z - r_0 \right).
\label{masterbar}
\ea
We have chosen to define strong phases so that $T$ is a real positive quantity.
Notice that $a_c + a_n = T e^{- i \alpha}$,
and
$q/p\, (\bar a_c + \bar a_n) = e^{2 i \theta_d} T e^{i \alpha}$.
Therefore,
the imaginary (real) part of the ratio of these quantities,
measures the sine (cosine) of the phase in Eq.~(\ref{phase-measured}).

%%%%% This paragraph should probably go somewhere else.
%%%%% But I have no idea where. I'll leave it here for now.
If $a_{00}$ and $\bar a_{00}$ are indeed color suppressed,
then $z$ and $r_0$ will be smaller than $1$, $z_1$ and $r_1$.
In that case,
the $CP$-violating difference
\be
\frac{|a_c|^2 - |\bar a_c|^2}{2 T^2}
=
- 2 \mbox{Re}\,  r_0 + 2 \mbox{Re} (z r_0^\ast)
\ee
will also be small.
%%%%%%%%%%%%%%%

In connection with  Eqs.~(\ref{decayamp:+-0}) and (\ref{decayamp:cdn}),
there is no advantage of
one parametrization with respect to another.
We could equally well have chosen to write the amplitudes in a
new set of basis functions $\{ f_1, f_2, f_3 \}$,
given by
\be
f_1 = \frac{f_+ + f_- + 2 f_0}{6},
\ \ \ \ \ 
f_2 = \frac{f_+ - f_-}{2},
\ \ \ \ \ 
f_3 = \frac{f_+ + f_- - f_0}{3}.
\ee
In this basis, the amplitudes would become
\ba
A 
& = &
f_1 \left( a_{c} + a_n \right) +
f_2 a_{d} + f_3 \left( a_{c} - a_n/2 \right)
\nonumber\\
\bar A
& = &
f_1 \left( \bar a_{c} + \bar a_n \right) +
f_2 \bar a_{d} + f_3 \left( \bar a_{c} - \bar a_n/2 \right).
\label{decayamp:123}
\ea
Eqs.~(\ref{decayamp:+-0}) highlight the intermediate $\rho \pi$
states;
Eqs.~(\ref{decayamp:cdn}) highlight the $CP$ structure of the intermediate 
charged $\rho \pi$ states, but still treat the intermediate 
$\rho^0 \pi^0$ separately;
and Eqs.~(\ref{decayamp:123}) highlight the extraction of 
$a_c + a_n = T e^{- i \alpha}$.
Although each basis has its pedagogical advantages,
they have no experimental significance.
What one does experimentally is a maximum likelihood
fit of  the variables $T$, $z_1$, $r_1$, $z$, and $r_0$ to
all the data in the Dalitz plots.

\subsection{Observables in the decay rates}

In the $B_d$ system we have $|q/p| \sim 1$ and $\Delta \Gamma \ll \Gamma$.
Therefore,
\ba
\Gamma[B^0 \rightarrow f] + \Gamma[\overline{B^0} \rightarrow f]
& \propto & |A_f|^2 + |\bar A_f|^2,
\label{untaggeddecayrate}
\\
\Gamma[B^0 \rightarrow f] - \Gamma[\overline{B^0} \rightarrow f]
& \propto & \left( |A_f|^2 - |\bar A_f|^2\right) \cos{\Delta m t}
-
2 \mbox{Im} \left( \frac{q}{p} \bar A_f A_f^\ast \right) \sin{\Delta m t}.
\label{difference}
\ea
One may rewrite these expressions using 
$\lambda_f = q \bar A_f / (p A_f)$.\footnote{Notice,
however,
that some authors use the opposite sign
convention for $q/p$ and, thus, for $\lambda_f$.}
The three terms in Eqs.~(\ref{untaggeddecayrate}) and (\ref{difference})
become $1 + |\lambda_f|^2$,
$1 - |\lambda_f|^2$,
and $\mbox{Im} \lambda_f$,
respectively.
The untagged decay rate in Eq.~(\ref{untaggeddecayrate})
probes only the combination $1 + |\lambda_f|^2$,
regardless of whether these measurements are time-dependent 
or time-integrated.
Due to the anti-symmetric nature of the $B^0-\overline{B^0}$
pairs produced at the $\Upsilon(4s)$,
tagged, time-integrated measurements performed at facilities working on this
resonance can probe $1 + |\lambda_f|^2$ and $1 - |\lambda_f|^2$,
but not $\mbox{Im} \lambda_f$.

If $f$ is a CP eigenstate and the decay amplitudes
$A_f$ and $\bar A_f$ are defined in such a way that the rates
$|A_f|^2$ and $|\bar A_f|^2$ include all the phase space integrations,
then $1 + |\lambda_f|^2$ is $CP$-even,
while
$1 - |\lambda_f|^2$ and $\mbox{Im} \lambda_f$ are $CP$-odd.
Under these conditions,
the ratio of Eq.~(\ref{difference}) to Eq.~(\ref{untaggeddecayrate})
is the famous $CP$-violating asymmetry.
It is sometimes stated that one can only probe 
$\mbox{Im} \lambda_f/(1 + |\lambda_f|^2)$.
While this is true if one looks only at the
$CP$-violating asymmetry,
one can see from Eqs.~(\ref{untaggeddecayrate}) and (\ref{difference})
that,
in principle,
there is enough information in the two decay rates for a clean
determination of $\mbox{Im} \lambda_f$.
The caveat is that disentangling $\mbox{Im} \lambda_f$
from $1 \pm |\lambda_f|^2$ requires that all these quantities be
affected by the same normalization.
This may not be true once the experimental cuts,
in particular possible cuts on the time $t$,
are folded into the analysis.

Using Eqs.~(\ref{decayamp:cdn}) we find
\be
|A|^2 \pm |\bar A|^2
= 
\sum_{i}
\left( |a_i|^2 \pm |\bar a_i|^2
\right) |f_i|^2
+
2
\sum_{i < j}
\mbox{Re} \left[ f_i f_j^\ast 
\left( a_i a_j^\ast \pm \bar a_i \bar a_j^\ast \right) 
\right],
\ee
where $i$ and $j$ can take the values $c$, $d$, and $n$.
The notation $i < j$ means that the $(i,j)$ pairs are not repeated.
Untagged decays probe the observables corresponding to
the $+$ sign.
Tagged,
time-integrated decays probe the observables with both signs
(or, what is the same, they measure $|A|^2$ and $|\bar A|^2$ separately).
Since the functions $f_i$ are quite distinct in their Dalitz plot
structure one can treat the coefficient of each different pair $f_i f_j^*$
as a separate observable.\footnote{Note,
however,
that the errors in the extraction of these various observables
are correlated.
Thus,
our observables,
while distinct are not technically independent.
A discussion of error correlations is beyond the scope of this paper,
but will of course be important in the application of this approach to data.}
We may define the nine untagged observables by
\be
\frac{|A|^2 + |\bar A|^2}{2}
= 
T^2
\left[
\sum_{i}
U_{ii}\, |f_i|^2
+
2
\sum_{i < j}
U_{ij}^R\, \mbox{Re} \left( f_i f_j^\ast \right) 
-
2
\sum_{i < j}
U_{ij}^I\, \mbox{Im} \left( f_i f_j^\ast \right)
\right],
\ee
where
\ba
U_{ii} &=& \frac{|a_i|^2 + |\bar a_i|^2}{2 T^2},
\nonumber\\
U^R_{ij} &=& 
\frac{\mbox{Re} \left( a_i a_j^\ast + \bar a_i \bar a_j^\ast \right)}{2 T^2}
\ \ \ \ (i \neq j),
\nonumber\\
U^I_{ij} &=& 
\frac{\mbox{Im} \left( a_i a_j^\ast + \bar a_i \bar a_j^\ast \right)}{2 T^2}
\ \ \ \ (i \neq j).
\label{U-observables:definitions}
\ea
We will refer to these observables generically as the $U$-observables.
We may define similar quantities with $U$ replaced by $D$
to describe the corresponding observables obtained for the
difference $|A|^2 - |\bar A|^2$.
These $D$-observables are obtained from 
Eqs.~(\ref{U-observables:definitions}) by replacing the
`+' signs with `-' signs.

Using Eqs.~(\ref{master}) and (\ref{masterbar}),
the untagged observables are
\ba
U_{cc} = \frac{|a_c|^2 + |\bar a_c|^2}{2 T^2}
& = &
|1-z|^2 + |r_0|^2,
\label{Ucc}
\\
U_{dd} = \frac{|a_d|^2 + |\bar a_d|^2}{2 T^2}
& = &
|z_1|^2 + |r_1|^2,
\label{Udd}
\\
U^R_{cd} + i U^I_{cd}
= \frac{a_c a_d^\ast + \bar a_c \bar{a}_d^\ast}{2 T^2}
& = &
r_1^\ast - z r_1^\ast - r_0 z_1^\ast,
\label{Ucd}
\\
U^R_{cn} + i U^I_{cn}
= \frac{a_c a_n^\ast + \bar a_c \bar{a}_n^\ast}{2 T^2}
& = &
 z^\ast - |z|^2 - |r_0|^2,
\label{Ucn}
\\
U^R_{dn} + i U^I_{dn}
= \frac{a_d a_n^\ast + \bar a_d \bar{a}_n^\ast}{2 T^2}
& = &
z_1 r_0^\ast + r_1 z^\ast,
\label{Udn}
\\
U_{nn} = \frac{|a_n|^2 + |\bar a_n|^2}{2 T^2}
& = &
|z|^2 + |r_0|^2.
\label{Unn}
\ea
Thus the $U_i$ contain nine different functions of the four complex
parameters $z_i, r_i$.
The requirement that the combination
$U_{cc} + 2 U^R_{cn} + U_{nn}$ equals $1$ yields
the overall normalization $T$.

Similarly,
the tagged,
time-integrated measurements will provide the additional observables
\ba
D_{cc} = \frac{|a_c|^2 - |\bar a_c|^2}{2 T^2}
& = &
- 2 \mbox{Re}\,  r_0 + 2 \mbox{Re} (z r_0^\ast) ,
\label{Dcc}
\\
D_{dd} = \frac{|a_d|^2 - |\bar a_d|^2}{2 T^2}
& = &
2 \mbox{Re} (z_1 r_1^\ast),
\label{Ddd}
\\
D^R_{cd} + i D^I_{cd}
= \frac{a_c a_d^\ast - \bar a_c \bar{a}_d^\ast}{2 T^2}
& = &
z_1^\ast - z z_1^\ast - r_0 r_1^\ast
\label{Dcd}
\\
D^R_{cn} + i D^I_{cn}
= \frac{a_c a_n^\ast - \bar a_c \bar{a}_n^\ast}{2 T^2}
& = &
r_0^\ast - 2 \mbox{Re} (z r_0^\ast),
\label{Dcn}
\\
D^R_{dn} + i D^I_{dn}
= \frac{a_d a_n^\ast - \bar a_d \bar{a}_n^\ast}{2 T^2}
& = &
z_1 z^\ast + r_1 r_0^\ast,
\label{Ddn}
\\
D_{nn} = \frac{|a_n|^2 - |\bar a_n|^2}{2 T^2}
& = &
2 \mbox{Re} (z r_0^\ast).
\label{Dnn}
\ea
Note that $D_{cc} + 2 D^R_{cn} + D_{nn} = 0$;
thus only eight different combinations of
$T$, $z_1$, $r_1$, $z$, and $r_0$ get fixed by measurements
of the $D$-observables.

Much can be learned about the interplay between the
Dalitz plot analysis and $CP$-violation by looking at
Eqs.~(\ref{Ucc}) through (\ref{Dnn}).
From the definitions of $r_0$ and $r_1$ in
Eqs.~(\ref{define:rz}),
we know that these quantities involve the product of a penguin
contribution with $\sin \alpha$.
Any nonzero value for $r_0$ and/or $r_1$  signals direct $CP$-violation.
It is true that such quantities do not by themselves
allow us to determine the size of $CP$-violation,
because $\sin \alpha$ appears multiplied by an unknown parameter;
the theoretical calculation of this parameter is plagued by
large hadronic uncertainties.
(This is as expected for any observable probing direct $CP$-violation.)

As expected,
the quantities $D_{cc}$, $D_{dd}$,
$D^R_{cn}$ , $D^I_{cn}$, and $D_{nn}$  are $CP$-odd.
Surprisingly,
the quantities
$U^R_{cd}$, $U^I_{cd}$, $U^R_{dn}$, and $U^I_{dn}$
are also $CP$-odd,
despite the fact that they are obtained by looking for
untagged decays.
How does this come about?
The reason is that there is a source of sensitivity to $CP$-violation induced
in the Dalitz plot analysis by the fact that $\rho_+$ and $\rho_-$ 
are $CP$-conjugate of each other.
Said otherwise,
when one performs a $CP$-transformation on $f_+ a_{+-}$,
one obtains $ f_- \bar a_{-+}$ and not a quantity proportional to $f_+$.
As pointed out before,
this means that a $CP$ transformation on $a_d$ yields $- \bar{a}_d$.
Therefore,
the quantities linear in $a_d$ and $\bar a_d$ have peculiar
$CP$-properties;
$U^R_{cd}$, $U^I_{cd}$, $U^R_{dn}$, and $U^I_{dn}$ are $CP$-odd,
while
$D^R_{cd}$, $D^I_{cd}$, $D^R_{dn}$, and $D^I_{dn}$ are $CP$-even.

Additional observables are obtained in the tagged,
time-dependent decays,
which contain a $\sin \Delta m\, t$ term given by
\be
\mbox{Im} \left( \frac{q}{p} \bar A\,  A^\ast \right)
= 
T^2 \left[
\sum_i I_{ii} |f_i|^2 
+
\sum_{i<j} I^I_{ij}\, \mbox{Re} \left( f_i f_j^\ast \right)
+
\sum_{i<j} I^R_{ij}\, \mbox{Im} \left( f_i f_j^\ast \right)
\right].
\ee
Here
\ba
I_{ii} &=&
\mbox{Im} \left( \frac{q}{p} \bar a_i  a_i^\ast \right)/T^2
,
\nonumber\\
I^I_{ij} &=&
\mbox{Im}
\left[ \frac{q}{p} 
\left( \bar a_i  a_j^\ast + \bar a_j  a_i^\ast \right)
\right]/T^2
\ \ \ \ (i \neq j),
\nonumber\\
I^R_{ij} &=&
\mbox{Re}
\left[ \frac{q}{p} 
\left( \bar a_i  a_j^\ast - \bar a_j  a_i^\ast \right)
\right]/T^2
\ \ \ \ (i \neq j).
\ea
As before, $i$ and $j$ take the values $c$, $d$, and $n$,
and the notation $i<j$ means that no $(i,j)$ pair gets repeated in the sum.
As a result,
we have nine new observables.
These observables depend on $\sin{2(\alpha + \theta_d)}$
and $\cos{2(\alpha + \theta_d)}$,
with coefficients given in table~I.
%%%%%%%%%%%%%%%%%%%%%%%%%%%%%%%%%%%%%%%%%%%%%%%%%%%%%%%%
%
% 1) Put at the end of the file before sending out to PRD
%
% 2) Separate out figure captions before sending out to PRD
%
%
%%%%%%%%%%%%  TABLE 1 %%%%%%%%%%%%

We stress that the uncertainties associated with the exact shape
chosen for the $\rho$ resonances are likely to affect
the $U_{ii}$, $D_{ii}$, and $I_{ii}$ coefficients
less than they affect the coefficients
$U_{ij}$, $D_{ij}$, and $I_{ij}$ with $i \neq j$

\section{Analysis}

This section reviews what can be learned in the various experimental searches.
Eight of the ten parameters needed to extract
$\alpha$ can be fit with untagged decays alone.
This greatly increases the data sample that will be available 
for determining these parameters,
since at the $B$-factories tagging efficiencies are estimated to
be of order $0.3$ (or less).
Further,
many questions about backgrounds and the contributions of the other
resonances to the three pion Dalitz plot can also begin to be
answered using this larger data sample of untagged events;
though they must also be re-examined in the tagged data sample,
where non-$B$ background will presumably be reduced.
Fitting to tagged,
time-integrated events,
fixes one further parameter and gives eight additional measurements 
that depend on combinations of the eight parameters already fixed,
thus improving the precision of their determination.
Only the important $CP$-violating CKM-related parameter 
$\alpha+\theta_d$ remains to be fit to the tagged time-dependent
Dalitz plot data.
Our conclusion is that these preliminary steps can and should be
performed at both symmetric and asymmetric $B$-factories.

\subsection{Observables from untagged decays}

The observables $U_{cc}$ through $U_{nn}$ can be combined to yield
$T$, $z$, $r_1$, $|z_1|$, $|r_0|$ and $\arg(z_1 r_0^\ast)$.
Therefore,
the nine observables present in untagged decays,
$U_{cc}$ through $U_{nn}$,
allow us to measure eight quantities and give one mathematically
redundant piece of information,
which of course serves to further constrain that combination of observables.

The result of this analysis is that the large data sample of untagged decays
is extremely important for the final determination of $\alpha$
in the $B \rightarrow \rho \pi$ channels.
One may use untagged decays to measure all relevant quantities except one
angle---the angle between $z_1$ (or $r_0$) and $z$ (or $r_1$)---and 
the $CP$-violating phase $\alpha + \theta_d$.
Moreover,
one will be sensitive to direct $CP$-violation through $|r_0|$ and $|r_1|$.

\subsection{Observables from tagged time-integrated decays}

The subset of events corresponding to tagged,
time-integrated decays will provide measurements of the
additional observables $D_{cc}$ through $D_{nn}$.
Since $U_{nn}$ and $D_{nn}$ are expected to be small,
it is interesting to note that the remaining 16 $U$- and $D$-observables
are sufficient to fix the nine parameters $T$, $z_1$, $r_1$, $z$,
and $r_0$,
and give seven mathematically redundant pieces of information.
That means that one does not need to probe quantities quadratic in
$|a_n|$ and $|\bar a_n|$
(that is, one does not need to have an observable $\rho^0 \pi^0$
branching fraction)
in order to determine all the observables attainable with these measurements.
Should $U_{nn}$ and $D_{nn}$ be measured,
they will provide two further mathematically redundant
pieces of information.
All these nine parameters have model-independent information that,
like a branching fraction,
can be taken from one experiment and used as input in another.
Therefore,
these experiments can and should be performed both at
CLEO and at the asymmetric $B$-factories.
The advantage of this is that fewer parameters remain to
be determined from the relatively small time-dependent data sample.

\subsection{Observables from tagged time-dependent decays}

Once a data sample of tagged, time-dependent events is available,
their rates contain a $\sin \Delta m\, t$ term,
allowing for a measurement of $I_{cc}$ through $I_{nn}$.
(One expects that, for some time to come,
this data sample will be small compared to the data samples discussed above.) 
The $I$-observables depend on $\sin{2(\alpha + \theta_d)}$
and $\cos{2(\alpha + \theta_d)}$,
with coefficients given in table~I.
We have already seen that the combination of untagged and
tagged time-independent decays yield $T$, $z_1$, $r_1$, $z$, and $r_0$.
These must now be combined with the observables in $I_{cc}$
through $I_{nn}$.
A combination of any pair of these $I$-observables yields
$\sin{2(\alpha + \theta_d)}$ and $\cos{2(\alpha + \theta_d)}$
independently.
A fit to all the observables determines $2(\alpha + \theta_d)$
(up to discrete ambiguities) and provides additional mathematically
redundant pieces of information.

We should point out that,
as expected,
one cannot extract $\alpha + \theta_d$
unless some information is known about quantities linear
in $a_n$ and $\bar a_n$.
However,
since these amplitudes into neutral $\rho^0 \pi^0$ might
be color suppressed,
it is interesting to ask what one may learn while only bounds,
rather than measurements,
on $|a_n|$ and $|\bar a_n|$ are known.
In section~\ref{usefulbound} we will show that {\em bounds} on 
these quantities can be combined with {\em measurements}
of quantities involving the charged $\rho^\pm \pi^\mp$
channels in order to determine
$\alpha + \theta_d$, 
up to an error that decreases with  $|a_n|$ and $|\bar a_n|$.

\section{Two useful bounds}
\label{usefulbound}

In this section we suppose that the bands in the Dalitz plot
corresponding to $B \rightarrow \rho^\pm \pi^\mp$ have been
measured while only bounds on the pieces linear and quadratic
in $a_n$ and $\bar a_n$ are known.
We will show that one may still find bounds on 
$\alpha + \theta_d$,
with an error that decreases as $|a_n|$ and $| \bar a_n|$ decrease.

\subsection{One useful bound}

As we have seen before,
a measurement of quantities which are independent of $a_n$ and
$\bar a_n$ is not enough for a determination of $\alpha + \theta_d$.
In particular,
we can see from Eqs.~(\ref{Ucc}),  (\ref{Dcc}) and
the first entry in Table~1
that measuring the parameters $U_{cc}$, $D_{cc}$, and $I_{cc}$,
which refer only to the decays into $\rho^\pm \pi^\mp$,
is enough to
determine\footnote{Notice that
$\frac{I_{cc}}{\sqrt{U_{cc}^2 - D_{cc}^2}}
\sim \frac{I_{cc}}{U_{cc}} \left( 1 + \frac{D_{cc}^2}{2 U_{cc}^2} \right)$.
Therefore,
measurements of $I_{cc}$ and $U_{cc}$ determine 
$\sin{ \left[ 2 ( \alpha + \theta_d + \delta_\alpha) \right]}$,
up to an error that is of second order in $D_{cc}$
(second order in $r_0$).}
\be
\frac{I_{cc}}{\sqrt{U_{cc}^2 - D_{cc}^2}}
=
\frac{\mbox{Im} \left( \frac{q}{p} \frac{\bar a_c}{a_c} \right)}{
\left| \frac{q}{p} \frac{\bar a_c}{a_c} \right| }
= \sin{ \left[ 2 ( \alpha + \theta_d + \delta_\alpha) \right]},
\label{restricted-lambda}
\ee
where we have defined
\be
2 \delta_\alpha = \arg{\left( \frac{1-z+r_0}{1-z-r_0} \right)}.
\label{delta-alpha}
\ee
Unfortunately,
$\delta_\alpha$ is neither zero nor calculable.

However,
as we will now show,
bounds on $|a_n|$ and $|\bar a_n|$ are enough to constrain
$\delta_\alpha$.
As these bounds decrease,
the deviation of Eq.~(\ref{restricted-lambda}) from 
$\sin{2(\alpha + \theta_d)}$ also decreases.
\begin{figure}
\centerline{\psfig{figure=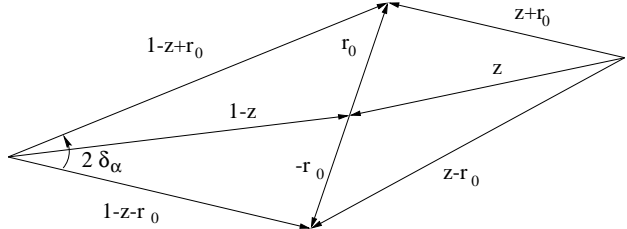,height=1.2in}}
\caption{Triangle construction for $1-z \pm r_0$ and $z \pm r_0$.
These complex numbers are related to
$\bar a_c$, $a_c$, $a_n$, and $\bar a_n$,
respectively.  
\label{fig:1}}
\end{figure}
To prove this,
we use Fig.~1 to derive
\be
|2 r_0|^2 = |1-z+r_0|^2 + |1-z-r_0|^2
- 2 |1-z+r_0|\, |1-z-r_0|\, \cos{(2 \delta_\alpha)},
\label{forthebound-1}
\ee
and
\be
|2 r_0| \leq |z+r_0| + |z-r_0|.
\label{forthebound-2}
\ee
Using these equations and the expressions
for $|a_n|$, $|\bar a_n|$, $|a_c|$, and $|\bar a_c|$
in Eqs.~(\ref{master}) and (\ref{masterbar}),
we find
\be
\sin^2 \delta_\alpha
\leq
\frac{\left( |a_n| + |\bar a_n| \right)^2
- \left( |a_c| - |\bar a_c| \right)^2}{
4 |a_c \bar a_c|}.
\label{eq:thebound1}
\ee
Notice that,
since we assume $|a_c|$ and $|\bar a_c|$ to be known,
we only need to know bounds on quantities linear in
$a_n$ and $\bar a_n$ in order to get bounds on
$|a_n|$ and $|\bar a_n|$.
The observables quadratic in $a_n$ and $\bar a_n$ are not needed.
We have thus proved that one can determine $(\alpha + \theta_d)$
by {\em measuring} only quantities from charged $\rho^\pm \pi^\mp$
final states,
up to an error that decreases as the {\em bounds} on the
color suppressed amplitudes $B \rightarrow \rho^0 \pi^0$
decrease.

It is interesting to compare the bound in Eq.~(\ref{eq:thebound1}),
with similar bounds obtained previously for the $B \rightarrow \pi \pi$
decays \cite{grossmanquinn,charles,pirjol}.
We note a similarity between the parametrization of the $B \rightarrow \pi \pi$
decays and the parametrization of the $B \rightarrow \rho \pi$ decays,
related through
\ba
a_n & \leftrightarrow &
2\,  A\left(B^0 \rightarrow \pi^0 \pi^0 \right),
\nonumber\\
a_c & \leftrightarrow &
\sqrt{2}\, A\left(B^0 \rightarrow \pi^+ \pi^- \right),
\ea
and similarly for the complex conjugated channels.
After these substitutions and some straightforward algebra we can
write a bound similar to that in Eq.~(\ref{eq:thebound1}) as
\be
\cos{(2 \delta_{\pi \pi})}
\geq
\frac{1}{\sqrt{1 - a_{\rm dir}^2}}
\left[ 1 - 2
\frac{\left( \sqrt{{\cal B} (B^0 \rightarrow \pi^0 \pi^0)} +
\sqrt{{\cal B} (\overline{B^0} \rightarrow \pi^0 \pi^0)} \right)^2}{
{\cal B} (B^0 \rightarrow \pi^+ \pi^-) +
{\cal B} (\overline{B^0} \rightarrow \pi^+ \pi^-)}
\right],
\label{incosineform}
\ee
where
\be
a_{\rm dir} = \frac{{\cal B} (B^0 \rightarrow \pi^+ \pi^-)
-{\cal B} (\overline{B^0} \rightarrow \pi^+ \pi^-)}{
{\cal B} (B^0 \rightarrow \pi^+ \pi^-)
+ {\cal B} (\overline{B^0} \rightarrow \pi^+ \pi^-)}.
\ee
If there is only data on untagged $B \rightarrow \pi^0 \pi^0$
decays,
we may still obtain a bound by substituting the squared
quantity on the right hand side (RHS) of Eq.~(\ref{incosineform}) by
$2 {\cal B} (B^0 \rightarrow \pi^0 \pi^0) +
2 {\cal B} (\overline{B^0} \rightarrow \pi^0 \pi^0)$,
a bound previously derived by Charles \cite{charles}.
If, in addition,
there is no data on $a_{\rm dir}$,
then we may obtain a weaker bound by setting $a_{\rm dir}$
to one on the RHS of Eq.~(\ref{incosineform})
\cite{grossmanquinn,charles,pirjol}.
In this form,
the bound depends only on untagged data and it is 
related to a bound obtained earlier by Grossman and Quinn
\cite{grossmanquinn} by using
$B^\pm \rightarrow \pi^+ \pi^0$ decays
instead of $B \rightarrow \pi^+ \pi^-$ decays
on the RHS of Eq.~(\ref{incosineform}).

\subsection{A bound from interference effects}

One may find a much cleaner bound by rewriting
Eq.~(\ref{delta-alpha}) in the form
\be
2 \delta_\alpha = \arg{\frac{1+ x}{1-x}},
\ee
where
\be
x = \frac{r_0}{1-z} = - \frac{D_{cc} + D^R_{cn} + i D^I_{cn}}{
U_{cc} + U^R_{cn} + i U^I_{cn}}.
\label{boundonx}
\ee
We find
\be
\tan{(2 \delta_\alpha)}
= \frac{2 \mbox{Im} x}{1 - |x|^2}.
\label{helenbound}
\ee
If the quantities $z$ and $r_0$ are of order one,
then we expect to be able to measure them.
The case of interest here is when $z \ll 1$,
$r_0 \ll 1$,
and the best we can do is place bounds on quantities linear in these
variables, while we expect to be able to measure $U_{cc}$ and $I_{cc}$.
In this case,
we can constrain the allowed values for $\tan{(2 \delta_\alpha)}$
by combining the measurement of $U_{cc}$ with the bounds
on $U^R_{cn}$, $U^I_{cn}$, $D^R_{cn}$ and $D^I_{cn}$.
(In effect,
to gain any information,
we need the bounds on the magnitudes of the latter quantities
to be smaller than the value for  $U_{cc}$.)

Notice that Eq.~(\ref{boundonx}) involves
$U^R_{cn}$, $U^I_{cn}$, $D^R_{cn}$ and $D^I_{cn}$
which are linear in $a_n$ and $\bar a_n$.
That means that here one is using the interference between the tails of two
different $\rho \pi$ channels.
This has two consequences.
Firstly,
this bound,
which is extremely powerful,
has no analogue for the $B \rightarrow \pi \pi$ decays.
Secondly,
since the bound depends on the interference effects,
it may be sensitive to the assumptions
mentioned above about the exact shape of the $\rho$-resonances.

\section{Some experimental issues}

This analysis has taken a purely formal approach and has not
evaluated all the relevant experimental questions.
The quantities which we call ``mathematically redundant''
are actually additional data samples that contribute to
fixing the parameters.
The parameters we define are all intrinsic to the physical process
and hence are,
like branching fractions,
expected to be the same in any experiment.
One issue that will be important experimentally is that the errors on the
various parameters are highly correlated and must be treated
correctly in establishing bounds such as those from Eq.~(\ref{helenbound})
and the allowed range for $\alpha$.
Moreover,
backgrounds,
efficiencies,
and cuts will differ in the different data samples and must be
investigated separately in each case.
A good knowledge of the variations in efficiencies across the
Dalitz plot is also essential for this analysis.
The extraction of the parameter $T$ is sensitive to the
knowledge of overall efficiencies;
it may be better to simply define a
$T_{eff}$ measured separately for each data sample than to depend on the
accuracy with which the overall efficiency and luminosity
for each data sample is known.
Other caveats have been mentioned,
such as the need to explore the sensitivity of the results
to reasonable changes in the assumed $\rho$-shape parameterization,
and the recognition of the possible large backgrounds in the untagged sample.
All of these issues and many more will only be settled by examining the
data.\footnote{One can get some idea of the impact of some of them
by simulations based on model values for the parameters.
Such studies are in progress \cite{Cahn}.}
None of them appear to us to invalidate the expectation that it will be
very valuable,
at least in the early years of study of these modes,
to use the parameters determined from the time-integrated experiments
when studying the $CP$ violating effects in the time-dependent data.

\section{Conclusions}
\label{sec:conclusions}

The $B \rightarrow \rho \pi$ decays are described by ten parameters,
{\it c.f.\/} Eqs.~(\ref{master}) and (\ref{masterbar}).
We have shown that untagged data can be
used to extract eight of these ten parameters and tagged time-integrated
data allows evaluation of one further parameter.
This leaves only the one $CP$-violating parameter $\alpha$
($\alpha + \theta_d$ if new physics contributes an extra phase $\theta_d$
to the mixing)
to be determined from time-dependent data.
The parameters in question are  defined in an experiment-independent
fashion and hence the values measured in a time-integrating experiment,
such as can be pursued at CLEO,
can be used as input to fits of the time-dependent data sample
-- thereby possibly allowing a measurement of the parameter
$\alpha + \theta_d$ earlier than could be achieved without this input. 

We have also shown that,
if the neutral $\rho$ contributions are small,
then,
prior to the time when the statistics are sufficient to provide a measurement
of these effects,
bounds on their contribution will allow bounds on the shift
of the angle measured from the rates and interference of the two charged
$\rho$ channels from the true value $\alpha$.

\acknowledgments

We are indebted to Y.\ Gao for discussions concerning the
$B \rightarrow \rho \pi$ measurements at CLEO,
and to Y.\ Grossman for discussions regarding the bounds
on $\delta_\alpha$ and for reading this manuscript.
This work is supported by the Department of Energy 
under contract DE-AC03-76SF00515.
The work of J.\ P.\ S.\ is supported in part by Fulbright,
Instituto Cam\~oes, and by the Portuguese FCT, under grant
PRAXIS XXI/BPD/20129/99	and contract CERN/S/FIS/1214/98.

\appendix

%%%%%%%%%%%%%%%%%%%%%%%%%%%%%%%%%%%%%%%%%%%%%%%%%%%%%%%%%%%%%%%%
%			APPENDIX A
%%%%%%%%%%%%%%%%%%%%%%%%%%%%%%%%%%%%%%%%%%%%%%%%%%%%%%%%%%%%%%%%
%\section{Parametrizing the decay amplitudes}
%
% Would use this form without the * if there are more than one appendix
%%%%%%%%%%%%%%%%%%%%%%%%

\section*{Parametrizing the decay amplitudes}
\label{app:A}

In this appendix we parametrize the $B^0 \rightarrow \pi^+ \pi^- \pi^0$
decay amplitudes by breaking them into $CP$-even and $CP$-odd components.
This will allow us to see more clearly what quantities may
be measured with the various types of experiments.
We may decompose the isospin amplitudes into tree-level and
penguin contributions as
\ba
A_{3/2, 2} & = & - T_{3/2, 2} e^{i \gamma},
\nonumber\\
A_{3/2, 1} & = & - T_{3/2, 1} e^{i \gamma},
\nonumber\\
A_{1/2, 1} & = & - T_{1/2, 1} e^{i \gamma} + P_{1/2, 1} e^{- i \beta},
\nonumber\\
A_{1/2, 0} & = & - T_{1/2, 0} e^{i \gamma} + P_{1/2, 0} e^{- i \beta},
\ea
where we have used the fact that $A_{3/2,1}$ and $A_{3/2,2}$
only receive contributions from the tree-level diagrams.
For convenience,
we have included an explicit minus sign in the definitions of
the tree-level amplitudes.
Substituting into Eqs.~(\ref{eq:clebsch}) and (\ref{define:acd}),
and dropping the $\Delta I = 5/2$ amplitude,
we find
\ba
e^{i \beta} a_c
&=&
T e^{- i \alpha}
\left[
\frac{1}{3} + \sqrt{\frac{2}{3}} \frac{T_{1/2, 0}}{T}
+ \sqrt{\frac{2}{3}} \frac{P_{1/2, 0}}{T} \cos{\alpha}
+ i \sqrt{\frac{2}{3}} \frac{P_{1/2, 0}}{T} \sin{\alpha}
\right],
\nonumber\\
e^{i \beta} a_d
&=&
T e^{- i \alpha}
\left[
\frac{T_{3/2, 1}}{T} + \frac{T_{1/2, 1}}{T}
+ \frac{P_{1/2, 1}}{T} \cos{\alpha}
+ i \frac{P_{1/2, 1}}{T} \sin{\alpha}
\right],
\nonumber\\
e^{i \beta} a_n
&=&
T e^{- i \alpha}
\left[
\frac{2}{3} - \sqrt{\frac{2}{3}} \frac{T_{1/2, 0}}{T}
- \sqrt{\frac{2}{3}} \frac{P_{1/2, 0}}{T} \cos{\alpha}
- i \sqrt{\frac{2}{3}} \frac{P_{1/2, 0}}{T} \sin{\alpha},
\right]
\ea
where we have defined $T = \sqrt{3} T_{3/2,2}$.

Let us define
\ba
z
& = &
\frac{2}{3} - \sqrt{\frac{2}{3}} \frac{T_{1/2, 0}}{T}
- \sqrt{\frac{2}{3}} \frac{P_{1/2, 0}}{T} \cos{\alpha},
\nonumber\\
z_1
& = &
\frac{T_{3/2, 1}}{T} + \frac{T_{1/2, 1}}{T}
+ \frac{P_{1/2, 1}}{T} \cos{\alpha},
\nonumber\\
r_0
& = &
- i \sqrt{\frac{2}{3}} \frac{P_{1/2, 0}}{T} \sin{\alpha},
\nonumber\\
r_1
& = &
i \frac{P_{1/2, 1}}{T} \sin{\alpha}.
\label{define:rz}
\ea
The parameter $z_1$ ($z$) contain the $CP$-even contributions
to the final state with isospin $I_f = 1$ ($I_f=0$ and also $I_f=2$),
while $r_1$ ($r_0$) contains the $CP$-odd penguin contributions to
the final state with isospin $I_f = 1$ ($I_f=0$). 
Using these definitions,
we arrive at 
\ba
e^{i \beta} a_c
& = &
T e^{- i \alpha}
\left( 1 - z - r_0 \right),
\nonumber\\
e^{i \beta} a_d
& = &
T e^{- i \alpha}
\left( z_1 + r_1 \right),
\nonumber\\
e^{i \beta} a_n
& = &
T e^{- i \alpha}
\left( z + r_0 \right).
\ea
Eqs.~(\ref{master}) are obtained from these by dropping the (irrelevant)
overall phase factor $e^{i \beta}$.
Similarly,
we may reach Eqs.~(\ref{masterbar})
by applying $CP$-conjugation,
multiplying by $q/p$,
and removing the same overall phase factor $e^{i \beta}$.

We may also parametrize the amplitudes involved
in the $B^\pm \rightarrow \rho^0 \pi^\pm \rightarrow \pi^+ \pi^- \pi^\pm$
and $B^\pm \rightarrow \rho^\pm \pi^0 \rightarrow \pi^\pm \pi^0 \pi^0$
decay chains as
\be
e^{\pm i \beta} A_{B^\pm \rightarrow \rho^0 \pi^\pm}
=
\frac{T e^{\mp i \alpha}}{\sqrt{2}}
\left( \frac{1}{2} - z_1 - \delta_1 \mp r_1\right),
\ee
and
\be
e^{\pm i \beta} A_{B^\pm \rightarrow \rho^\pm \pi^0}
=
\frac{T e^{\mp i \alpha}}{\sqrt{2}}
\left( \frac{1}{2} + z_1 + \delta_1 \pm r_1\right),
\ee
respectively,
where we have dropped the $\Delta I = 5/2$ amplitudes.
These decays involve the new complex parameter
\be
\delta_1 = - \frac{3}{2} \frac{T_{3/2, 1}}{T}.
\ee
Therefore,
the information in
$B^\pm \rightarrow \rho^0 \pi^\pm \rightarrow \pi^+ \pi^- \pi^\pm$
decays by itself does not help in constraining the
parameters involved in the decays of the neutral $B$'s into three pions;
these two rates merely allow a determination of the magnitude and
phase of $\delta_1$.
The decays from charged $B$ mesons are only useful if one can
measure both the
$B^\pm \rightarrow \rho^0 \pi^\pm \rightarrow \pi^+ \pi^- \pi^\pm$
and the experimentally challenging
$B^\pm \rightarrow \rho^\pm \pi^0 \rightarrow \pi^\pm \pi^0 \pi^0$
decay chains.
For example,
one might use the information from the decays of the neutral $B$
to get $T$, $r_1$, and $z_1$,
and the $B^\pm \rightarrow \rho^0 \pi^\pm$ decays to get $\delta_1$.
One would then be able to predict the rates for
the $B^\pm \rightarrow \rho^\pm \pi^0$ decays.
Even with the additional direct CP asymmetry measurements
in these channels,
we see that we have no way of extracting a measurement
of $\sin \alpha$.
As before, only the quantity $r_1$ appears.

%%%%%%%%%%%%%%%%%%%%%%%%%%%%%%

\vspace{2cm}

\begin{table}[t]
\centering
\begin{tabular}{|c|c|c|}
\hfil Observable\hfil& 
\hfil Coefficient of\hfil &\hfil Coefficient of\hfil 
\\
\hfil \hfil& 
\hfil $\sin{2(\alpha + \theta_d)}$\hfil &
\hfil $\cos{2(\alpha + \theta_d)}$\hfil
\\\hline
$I_{cc}$ & 
$|1-z|^2 - |r_0|^2$ &
$2 \mbox{Im} \left[ r_0 (1-z)^\ast \right]$ 
\\
$I_{dd}$ & 
$|r_1|^2 - |z_1|^2$ &
$2 \mbox{Im} \left[ r_1 z_1^\ast \right]$ 
\\
$I^I_{cd}$ & 
$2 \mbox{Re} \left[ r_1 (1-z)^\ast + z_1 r_0^\ast \right]$ &
$2 \mbox{Im} \left[ (1-z) z_1^\ast + r_0 r_1^\ast \right]$ 
\\
$I^R_{cd}$ & 
$2 \mbox{Im} \left[ r_1 (1-z)^\ast + z_1 r_0^\ast \right]$ &
$2 \mbox{Re} \left[ (1-z) z_1^\ast + r_0 r_1^\ast \right]$ 
\\\hline
$I^I_{cn}$ & 
$2 \mbox{Re} \left[ z - |z|^2 + |r_0|^2  \right]$ &
$2 \mbox{Im} \left[ (1 - 2 z) r_0^\ast \right]$ 
\\
$I^R_{cn}$ &
$2 \mbox{Im}\, z$ &
$2 \mbox{Re}\, r_0$
\\
$I^I_{dn}$ & 
$2 \mbox{Re} \left[ z r_1^\ast - r_0 z_1^\ast \right]$ &
$2 \mbox{Im} \left[ r_1 r_0^\ast - z_1 z^\ast \right]$ 
\\
$I^R_{dn}$ & 
$2 \mbox{Im} \left[ z r_1^\ast - r_0 z_1^\ast \right]$ &
$2 \mbox{Re} \left[ r_1 r_0^\ast - z_1 z^\ast \right]$ 
\\\hline
$I_{nn}$ & 
$|z|^2 - |r_0|^2$ &
$2 \mbox{Im} \left[ z r_0^\ast \right]$ 
\\
\end{tabular}
\caption{$I$-observables obtained from the $\sin{\Delta m\, t}$ term.
The table shows their dependence on $\sin{2(\alpha + \theta_d)}$
and $\cos{2(\alpha + \theta_d)}$.}
\end{table}
%%%%%%%%%%%%%%%%%%%%%%%%%%%%%%%%% 

\end{document}